\documentclass[twocolumn]{openjournal}

\usepackage{natbib}
\usepackage{hyperref}
\usepackage{graphicx} % Required for inserting images
\usepackage{amssymb}	% Extra maths symbols
\usepackage{amsmath}
\usepackage{xcolor}

\begin{document}

\shorttitle{Rubble-Pile Collisions}
\shortauthors{Guidos, Kolanz, Lazzati}

\title{Discrete element simulations of self-gravitating rubble pile collisions: the effects of non-uniform particle size and rotation.}

\author{Job Guidos, Lucas Kolanz, and Davide Lazzati}

\affil{Department of Physics, Oregon State University, 301
  Weniger Hall, Corvallis, OR 97331, USA} 

%%%%%%%%%%%%%%%%%%%%%%%%%%%%%%%%%%%%%%%%%%%%%%%%%%
\begin{abstract}
We present a new implementation of a soft-sphere discrete element code to simulate the dynamics of self-gravitating granular materials. The code is used to study the outcome of sub-sonic collisions between self-gravitating rubble piles with masses ranging from $\sim6\times10^{21}$ to $\sim6\times10^{22}$~g. These masses are representative of asteroids and planetesimals in the $\sim100$~km range. We simulate rubble piles composed of a range of particle sizes and analyze the collision outcomes focusing on the properties of the largest surviving fragment. We successfully test and validate the code against previous results. The results of our study show that rubble piles formed by collisions of two parent rubble piles do not maintain the same particle size distribution as their parents. Rubble piles formed in low velocity collisions are characterized by a larger fraction of large particles, while the largest fragments of high-velocity collisions show a decrease in their mean particle size. In both cases the effect is small, but could build up to a noticeable difference after multiple collisions. We ascribe this effect to the fact that large particles transmit most of the forces during the collisions. In addition, we find that the mass of the largest post-collision fragment depends on the rotation of the colliding rubble piles. This effect is especially noticeable when the pre-collision spin axes are parallel with each other and perpendicular to the relative velocity. This finding can be particularly relevant for meter to kilometer sized bodies
embedded in protostellar accretion disks, where viscous stresses can efficiently align the target and projectile spin axes.
\end{abstract}

%%%%%%%%%%%%%%%%%%%%%%%%%%%%%%%%%%%%%%%%%%%%%%%%%%
\section{Introduction} \label{sec:intro}

While most of the baryonic matter in the Universe is in gaseous and plasma phase within stars, galaxies, and the intergalactic medium \citep{Copi1995,Peroux2020}, solids do play many crucial roles in the structure and evolution of the Cosmos \citep{Draine2003}. Solid phase materials are commonly found in the interstellar medium (ISM) as dust grains \citep{Mathis1990,Weingartner2001,Calzetti2000} and in planetary systems as rocky planets and minor bodies, such as asteroids, comets, etc \citep{Lissauer1993,Chapman1978,Walsh2018,Spinard1987}. Dust grains are known to have an effect on the thermodynamics \citep{Spitzer1978} of the ISM and to be receptacles for interstellar chemistry \citep{Duley1981,Hasegawa1993}. Collisions between grains lead to the formation of larger objects that can grow to become asteroids, planetesimals, and eventually terrestrial planets \citep{Goldreich1973,Johansen2007,Blum2008,Wyatt2008}. Some of the solids in space are whole, single pieces, while others --- like dust conglomerates or some asteroids and planetesimals --- are clusters made of loosely bound individual units or particles \citep{Murdoch2015,Sanchez2016,Walsh2018,Hestroffer2019}. The larger clusters are held together mainly by gravity, while the small ones are dominated by contact forces. 

\begin{table*}[]
    \centering
    \begin{tabular}{c|l|c|c}
    Symbol & $\qquad \quad$ Description & Value(s) &  Units  \\ \hline
    $R_i$ & Radius of the $i$-th particle  & $[0.75,1.5,2.25]\times10^{6}$ & cm \\ 
    $m_i$ & Mass of the $i$-th particle  & $[0.48,3.8,12.9]\times10^{19}$ & g \\ 
    $\rho$ & Particle density & 2.7 & g/cm$^3$ \\
    $\vec{r}_i$ & Position vector of the $i$-th particle & & cm \\
    $r_{ij}$ & Distance between the $i$-th and $j$-th particles & & cm \\
    $\hat{r}_{ij}$ & Unit vector that points from the $j$-th to the $i$-th particles' positions & & \\
    $\vec{v}_i$ & Velocity vector of the $i$-th particle  & & cm/s    \\
    $\vec{\omega}_i$ & Angular velocity vector of the $i$-th particle  & & rad/s \\
    $\vec{F}_{\rm{G},ij}$ & Gravity force on particle $i$ due to particle $j$  & & dyne \\
    $\vec{F}_{\rm{elastic}}$ & Normal elastic force between two particles   & & dyne    \\
    $\vec{F}_{i,\rm{friction}}$ & The kinetic friction force on particle $i$  & & dyne    \\
    $\vec{F}_{i,\rm{rolling\,friction}}$ & The rolling friction force on particle $i$  & & dyne    \\
    $\vec{T}_{ij}$ & The torque on particle $i$  due to its interaction with particle $j$ & & dyne cm   \\
    $k$ & Spring constant & $10^{18}$ & dyne/cm \\
    $\mu_k$ & Coefficient of kinetic friction & 0.3 & \\
    $\mu_r$ & Coefficient of rolling friction & $10^{-3}$ & \\
    $\delta{t}$ & time step of the code & 0.04 & s \\
    $m_{\rm{p}}$ & mass of the projectile (the smaller cluster) & & g \\
    $m_{\rm{t}}$ & mass of the target (the larger cluster) & & g \\
    $m_{\rm{lf}}$ & mass of the largest fragment after collision & & g
 \end{tabular}
    \caption{List of symbols used and their meaning.}
    \label{tab:parms}
\end{table*}

In general, based on observational evidence and previous studies \citep{Wetherill1967,Dohnanyi1969,Chapman1978,Farinella1982,Davis1985,Michel2001,Benavidez2012,Leinhardt2012,Walsh2018}, the result of collisions between clusters is not always constructive. Here, by constructive we mean that the largest debris of the collision is more massive than the target cluster\footnote{Here and in the following we will call the projectile the cluster with the lower mass, while we will call the target the more massive one. Since our calculations are carried out in the center of mass frame, in which the net momentum is zero, the projectile also has the largest velocity. In case the two rubble piles have the same mass, one of the two will arbitrarily be the projectile and the other the target, if such distinction is needed.}. Although higher-velocity collisions are expected to have a higher likelihood of being destructive, the collision outcome is not simply predictable by the energy balance of the collision. Previous work \citep{Farinella1982,Love1996,Asphaug1998,Durda2004,Richardson2009} has indeed shown that even collisions with kinetic energy larger than the binding energy (in absolute value) can be constructive, resulting in a massive bound cluster surrounded by fast debris that carry most of the excess kinetic energy. Additional energy is dissipated in inelastic collisions that increase the thermal energy of the particles. \cite{Leinhardt2012}, for example, found that even collisions in which the kinetic energy of the center of mass (CoM) is three times higher than the binding energy of the gravitational potential of the clusters can be constructive. The high resilience of granular objects (or clusters) to destructive collisions can be ascribed to several properties that differentiate them from gaseous bodies. Clusters of hard particles obey the dynamics of granular materials. In these structures the pressure gradient is not nearly
as isotropic as it is in stars. Instead, the gravity is balanced through contact forces between particles. This results in force chains that propagate collision stresses unevenly among particles. In some configurations, some particles may not be part of any force chains \citep{Clark2012,Clark2014,Cheng2018}. Since the forces generated by a collision are propagated through the clusters via contact points, the inelasticity of the particles that compose the cluster allows them to absorb a fraction of the impact energy at every contact point \cite{Tanga2009,Ferrari2020}. This causes some of the impact energy to be converted into internal energy, increasing the thermodynamic temperature of the cluster particles. Finally, impacts on porous media create scattered shocks whose energy is confined by the voids, leading to lower levels of disruption \citep{Asphaug1998}.

In this paper we present a new code to study the outcome of subsonic collisions between clusters with masses in the asteroid and planetesimal range. We first present the code (Section~\ref{sec:code}) and compare its results with the outcome of previous studies (Section~\ref{sec:validation}). We then use the capabilities of our code to study the effect of collisions on the particle size distribution of the pre- and post-collision clusters, as well as the effect of rotation of the projectile and target clusters (Section~\ref{sec:results}). We also analyze the dynamics of the collisions to tease out which particle size is mostly involved in propagating forces during the impact.
We finally discuss our results and compare with previous studies (Section \ref{sec:discussion}).

\section{The DECCO Code} 
\label{sec:code}

In this section we present our soft-sphere, discrete element code DECCO (Discrete Element Cosmic COllision) that we used in the calculations presented in section~\ref{sec:results}. This type of simulation was originally proposed for granular dynamics by \cite{Cundall71} (see also \citealt{Cundall79}) and is often used to model granular physics systems (e.g., \citealt{Hart1988,Mahmood2016,Sanchez2011}). However, alternative methods exist, such as contact dynamics, quasi-statics, and event driven methods (see, e.g., \citealt{Radjai2013} for a review). Our code uses the 
soft-sphere method to compute the binary collision between particle pairs\footnote{Collisions between multiple particles are also considered as the cumulative effect of multiple binary collisions.}, and includes the effects of inelastic collisions as well as kinetic and rotational friction between particles. At the present stage of development, DECCO only includes gravity as the attractive force and is therefore best suited for collisions between macroscopic bodies, such as rubble-pile asteroids or planetesimals. Since DECCO is based on individual particle-particle interactions, it is applicable for impact velocities that are lower than the speed of sound in the particles. Hydrodynamic codes would be needed for faster collisions (see, e.g., \citealt{Michel2001,Leinhardt2012,Michel2020}). DECCO also calculates explicitly all binary interactions, yielding increased accuracy {with respect to approximate methods (e.g., tree codes and mean field approximation)} at the cost of $O(n^2)$ computational cost. The half-step Verlet method \citep{Verlet1967} is used to integrate the equations of motion of the particles. The equations of the method read:
\begin{eqnarray}
x[i+1]&=&x[i]+v[i]dt+\frac12 a[i]dt^2 \nonumber \\
v\left[i+\frac12\right]&=&v[i]+\frac12 a[i]dt \\
v[i+1] &=& v\left[i+\frac12\right] + \frac12 a[i+1]dt \nonumber
\end{eqnarray}
where $x[i]$ is position at time step $i$, $v[i]$ is velocity at time step $i$, and $a[i]$ is acceleration calculated at position $x[i]$.

\subsection{Free particles}

When there is no contact between two particles, i.e., the distance $r_{ij}$ from their centers is larger than the sum of the two particle radii ($r_{ij}>R_i+R_j$), the only acting force is gravitational attraction. In this case the force that the $i$-th particle feels because of the presence of the $j$-th particle is:
\begin{equation}
    \vec{F}_{ij}=\vec{F_G}_{,ij}=-G\frac{m_i m_j}{r_{ij}^2}\hat{r}_{ij}
\end{equation}
where 
\begin{equation}
    \hat{r}_{ij} = \frac{\vec{r}_i-\vec{r}_j}{r_{ij}}
\end{equation}
is the unit vector that points from the position of the $j$-th particle to the position of the $i$-th one.

\begin{figure}
\centering
\includegraphics[width=1\linewidth]{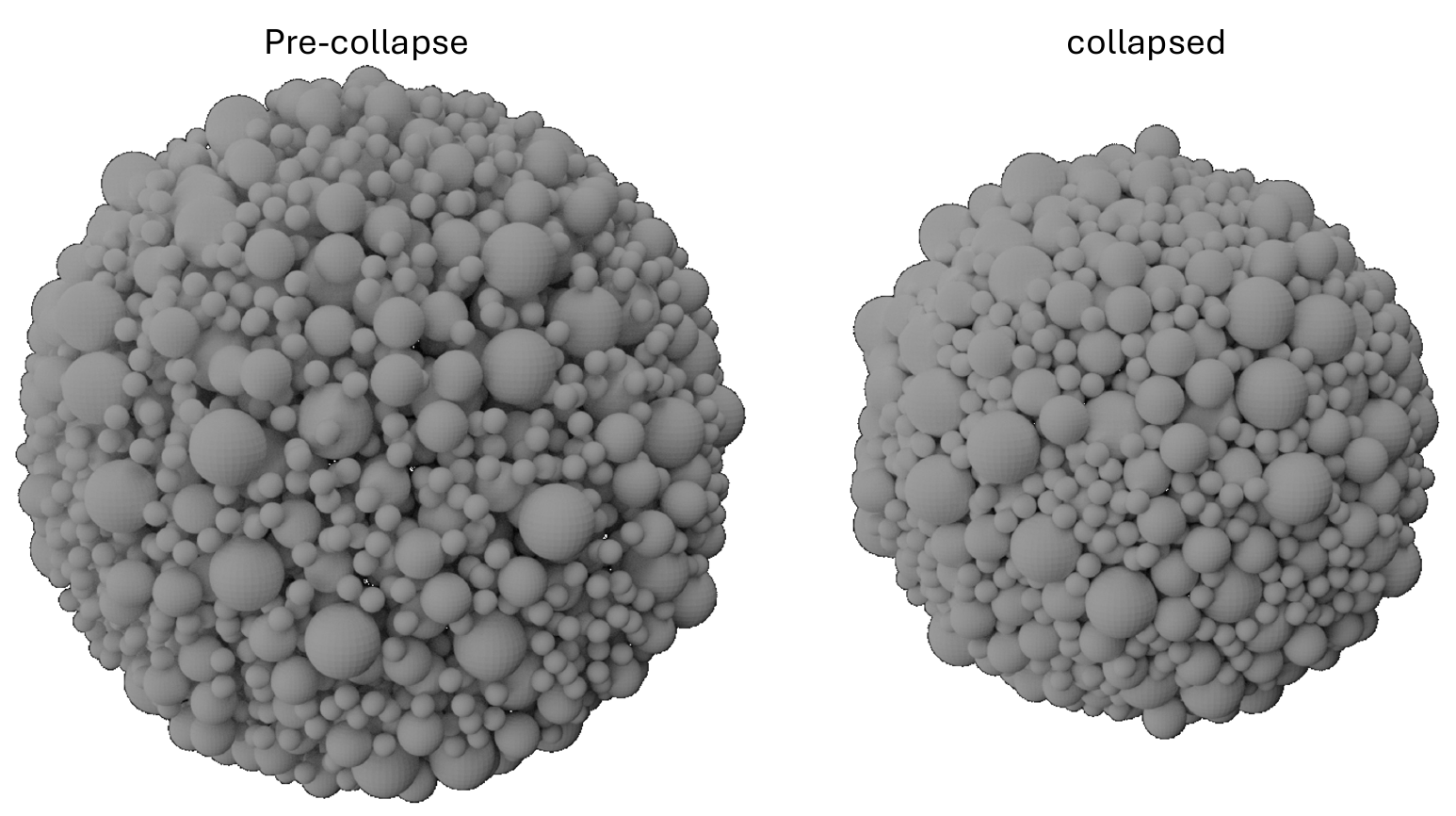}
\caption{Example of a pre-formed rubble-pile  (left) and the same after the collapse phase, when formation is completed (right).}
\label{fig:formation}
\vspace{3pt}
\end{figure}

\begin{figure*}
    \includegraphics[width=\textwidth]{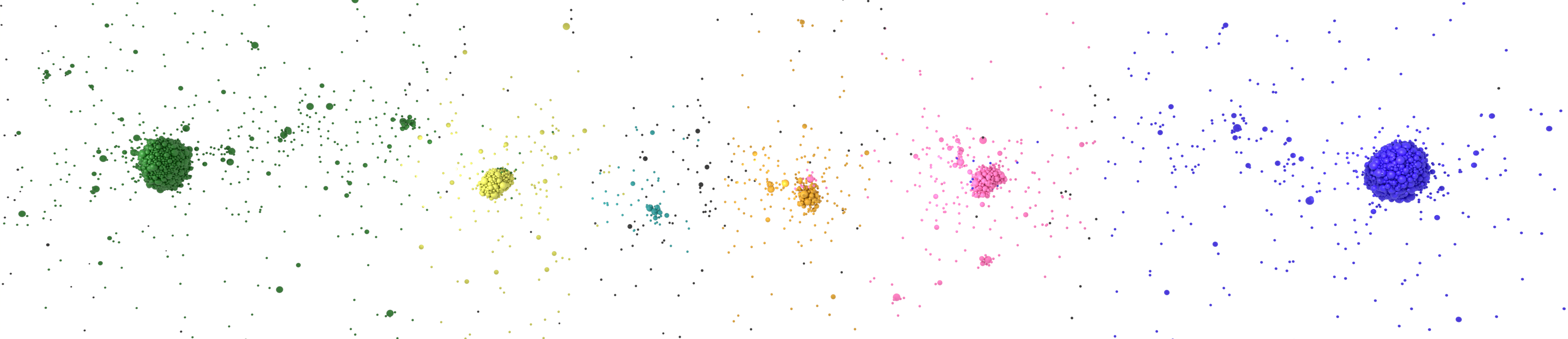}
    \caption{{Example outcome of the post-collision sub-cluster identification algorithm. The figure shows  the post-collision and re-collapse configuration from an $\eta=3$ collision between two identical rubble piles (5000 particles each). The area shown is where the largest debris are found. Particles belonging to dynamically bound sub-clusters with at least 50 particle members are color coded. Dark gray particles are free and unbound. There are six sub-clusters identified in the figure. The two biggest ones (blue and green) are almost identical due to the symmetry of the collision.}
    \label{fig:subclusters}}
\end{figure*}

\subsection{Radial contact forces}
The implementation of DECCO presented here includes the elastic repulsion between two touching particles
as the only radial contact force. To take into account the inelastic behavior of real materials, DECCO 
utilizes two different values of Hook's constant: a larger value $k_{\rm{in}}$ for the incoming trajectory (particles approaching each other) and a smaller value $k_{\rm{out}}$ for the outgoing trajectory.  This methodology is a simplified version of the linear hysteretic model (see, e.g., \citealt{Walton1986,Luding1998,Luding2007,An2007}) that ensures a constant coefficient of restitution. Writing conservation of energy for the inbound and outbound motion we have:
\begin{eqnarray}
    \frac12 m v_{\rm{in}}^2 &=& \frac12 k_{\rm{in}} \Delta x \nonumber\\ 
    \frac12 m v_{\rm{out}}^2 &=& \frac12 k_{\rm{out}} \Delta x \nonumber
\end{eqnarray}
where $\Delta x$ is the amount of compression of the spring. Taking the ratio of the two equations above one gets:
\begin{equation}
    \left(\frac{v_{\rm{in}}}{v_{\rm{out}}}\right)^2 = \frac{k_{\rm{in}}}{k_{\rm{out}}} \nonumber
\end{equation}
from which a constant coefficient of restitution $\epsilon$ is derived as:
\begin{equation}
    \epsilon=\frac{v_{\rm{out}}}{v_{\rm{in}}} =\sqrt{\frac{k_{\rm{out}}}{k_{\rm{in}}}}.
\end{equation}

The elastic force equation therefore reads as:
\begin{equation}
\vec{F}_{\rm{elastic}}=k\frac{R_i+R_j-r_{ij}}{2}\hat{r}_{ij}
\label{eq:spring}
\end{equation}
where $ r_{ij} $ is the distance between particles $i$ and $j$, $ R_i $ and $ R_j $ are the respective radii of particles $i$ and $j$, and the factor $\frac{R_i+R_j-r_{ij}}{2}$ measures the deformation of the spring, with the factor $\frac12$ taking into account that there are two springs in series \footnote{An alternative method to simulate the particle-particle bouncing force is by the use of a Hertzian spring  (see, e.g., \citealt{Landau1959}, or \citealt{Sanchez2011,Sanchez2012,Sanchez2014} for examples in the context of asteroids). For the kind of simulations discussed in this paper, we found that a Hertzian spring model causes increased computational cost without an appreciable difference in the outcome.}. The elastic constant $k$ is either $k_{\rm{in}}$ or $k_{\rm{out}}$, depending on whether the particles are approaching or receding, respectively.
When particles are colliding, this additional surface normal force is added to the gravitational force for the current timestep to simulate the collision `bounce'.

\subsection{Restitution}
As introduced above, Kinetic energy is removed from particles in collision by a coefficient of restitution directly applied through a change in the elasticity constant, $k$. This method allows for scale-free, constant restitution, which is particularly important since we aim at simulating objects made up of particles of different sizes.

\subsection{Friction}

To determine surface friction, two primary physical quantities must be derived. These are the relative surface velocity experienced by particle 1 from particle 2, and the normal force, $\vec{N}$, exerted during collision. The normal force of collision is the elastic repulsion force, $\vec{F}_{elastic}$ from Eq.~\ref{eq:spring}. The relative surface velocity that particle $i$ experiences from particle $j$ depends on both particles' angular velocity, as well as the component of their translational velocity tangent to their surfaces at the point of collision.

The surface velocity of particle $i$ at the point of contact with particle $j$ is found as:
\begin{equation}
\vec{v}_{i,surface}=  \hspace{-7mm}\underbrace{\vec{v}_i-(\vec{v}_i\cdot\vec{r}_{ij}) \hat{r}_{ij}}_{\text{vel. comp. tangent to surface\;\;\;\;\;\;\;\;}}\hspace{-7mm}+  
\underbrace{\vec{\omega}_i\times\frac{R_i}{R_i+R_j}\vec{r}_{ij}}_{\text{vel. due to rotation}}
\end{equation}
Where $ \vec{\omega_i} $ is angular velocity of particle $i$.
The relative surface velocity is then the difference between the surface velocities of the two particles:
\begin{equation}
\vec{v}_{i,surface,relative}=\vec{v}_{j,surface}-\vec{v}_{i,surface}
\end{equation}

The friction force experienced by particle $i$ is therefore
\begin{equation}
\vec{F}_{i,friction}=-\mu_k||\vec{F}_{\rm{elastic}}||\hat{v}_{i,surface,relative}
\end{equation}
where $\mu_k$ is the coefficient of friction. We emphasize that DECCO does not include a treatment for static friction nor twisting friction. As such, it does not satisfy Coulomb's friction criterion and therefore cannot be used to study long timescale stability of the clusters. In addition, the decrease in particle size due to compression was neglected. This can affect the calculation of the surface velocities as well as the torques between particles. Given the imposed limit on spring compression discussed below for code stability, these effects are expected to be minor.

In addition to sliding friction, the code includes rolling friction (e.g., \citealt{Lunding2008,Schwartz2012,Santos2020}), which is calculated as:
\begin{equation}
    \vec{F}_{i,rollingFriction}=-\mu_r||\vec{F}_{\rm{elastic}}||
    \frac{(\vec{\omega}_i-\vec{\omega}_j)\times \vec{r}_{ij}}{||(\vec{\omega}_i-\vec{\omega}_j)\times \vec{r}_{ij}||}
\end{equation}
where $\mu_r$ is the coefficient of rolling friction. It has to be noted that modeling rolling friction is a complicated task since any deviation from a pure spherical particle shape would affect their rolling behavior much more than their sliding behavior. To understand the reason, let us consider a small particle rolling on a big one. If the small particle  has a complex shape (without spherical or cylindrical symmetry), the rolling motion would cause the distance between the two centers of mass to change with time, a kind of motion that requires more energy than the rolling of spherical particles which does not involve any changes in distance (see also \citealt{Zhou2017} for further discussion).

The total friction force (sliding plus rolling) affects the center-of-mass linear velocities of the involved particles and contributes a torque that also affects their rotational velocities. The total torque on particle $i$ due to the contact with particle $j$ is then:
\begin{equation}
\vec{T}_{ij}=\frac{R_i}{R_i+R_j}\vec{r}_{ij}\times(\vec{F}_{i,friction}
+\vec{F}_{i,rollingFriction})
\end{equation}
and the equation for the evolution of angular velocity of particle $i$ is:
\begin{eqnarray}
&&\frac{2}{5} m_i R_i^2\frac{d\omega_i}{dt} = \nonumber \\
&=&\sum_{i\ne j}\frac{R_i}{R_i+R_j}\vec{r}_{ij}\times(\vec{F}_{i,friction}
+\vec{F}_{i,rollingFriction}) \qquad
\end{eqnarray}
where the inertia of the spherical particles has been assumed to be ${I}=\frac25 \, m_i R_i^2$. The friction force is equal and opposite for particle $j$, so the necessary information to move both interacting particles is known from a single calculation. The angular acceleration of a particle due to surface friction is found by dividing the torque on a particle by its moment of inertia.

\subsection{Code Stability}
\label{sec:stability}

\begin{figure*}
    \includegraphics[width=\textwidth]{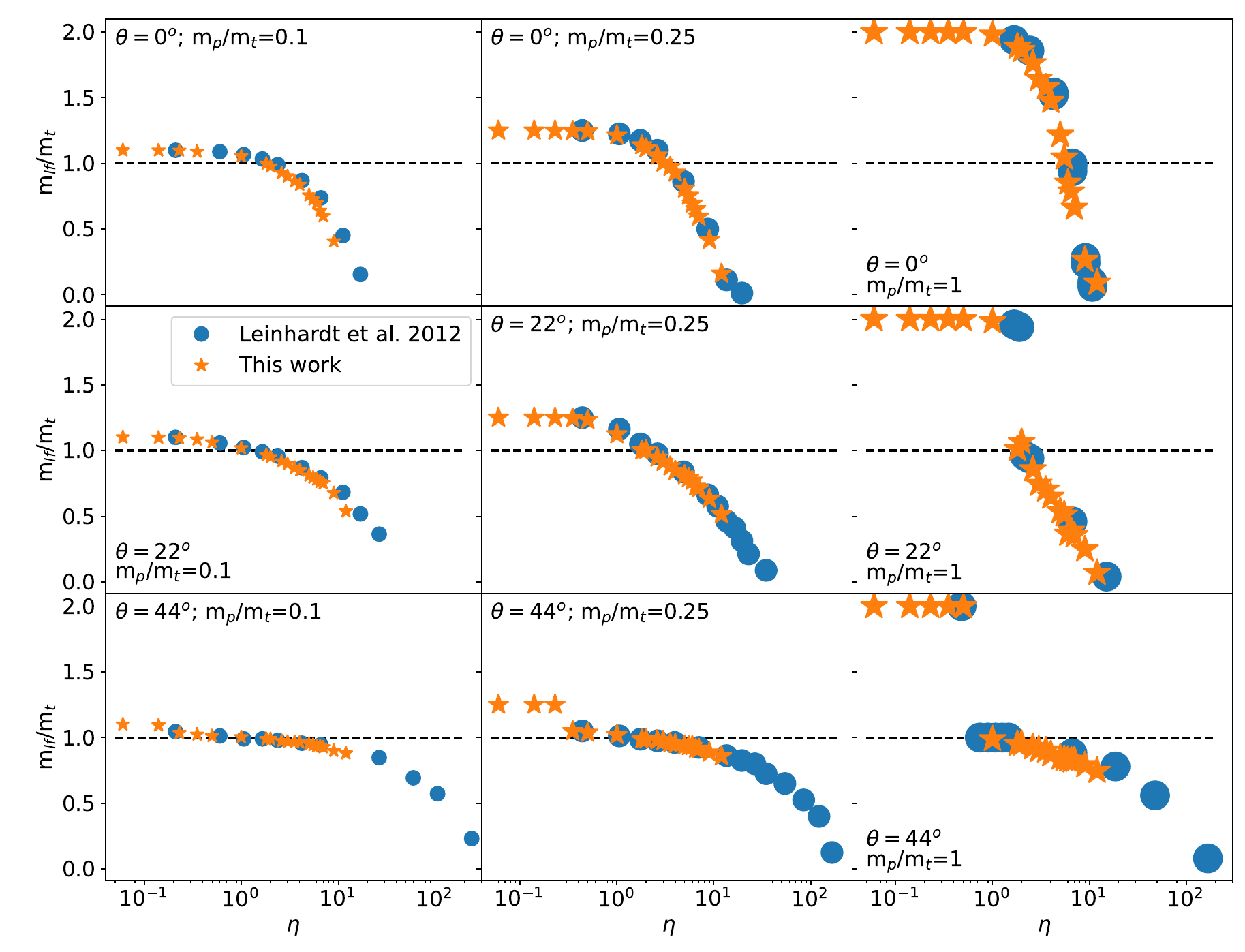}
    \caption{Comparison between our results and analogous simulations carried out by \cite{Leinhardt2012}. Each panel shows results for collision outcome between a projectile and a target with mass ratio $m_p/m_t$ for a specific collision angle $\theta$. Larger $\theta$ values are for grazing collisions, while $\theta=0^\circ$ represents a head-on collision. The {\it x-axis} shows the ratio of the total kinetic over total potential energies in the center of mass reference frame. The {\it y-axis}, instead, shows the ratio between the mass of the largest fragment ($m_{\rm{lf}}$) and the mass of the target cluster ($m_{\rm{t}}$). See also Figure~\ref{fig:cartoon} for a definition of the collision angle $\theta$.}
    \label{fig:comparaLei}
\end{figure*}

In order for the code to remain stable and conserve energy, a sufficiently large elastic constant and a sufficiently small time step need to be used.  We follow \cite{Herrmann1998} to derive constraints on the time step and elasticity constant. The requirements are that (i) the maximum compression $\Delta$ attainable in the simulation needs to be a small fraction $\epsilon_\Delta$ of the radius of the involved particles, and that (ii) at each time step $\delta{t}$ the movement of the fastest particle is only a small fraction $\epsilon_t$ of the maximum compression $\Delta$. For two particles $i$ and $j$ colliding with velocity $v_i$ and $v_j$ in their centers of mass, the compression can be written as:
\begin{equation}
\Delta_{ij}=\sqrt{\frac12\frac{m_iv_i^2}{k}\left(1+\frac{m_i}{m_j}\right)}
\end{equation}
In order to ensure stability (i.e., that particles cannot excessively penetrate each-other) we require:
\begin{equation}
    \Delta_{ij}\ll\frac{r_i+r_j}{2}
\end{equation}
Rearranging the various terms and using $m_i=4\pi/3 \, \rho r_i^3$ we obtain:
\begin{equation}
    k\gg2 \frac{\frac{4\pi}{3} \rho r_i^3 v_i^2 \left(1+\frac{m_i}{m_j}\right)}
    {r_i^2 \left(1+\frac{r_j}{r_i}\right)^2}
\end{equation}
where $\rho$ is the particle density (see Table~\ref{tab:parms} for a list of parameters used). If we assume that $i$ is the lighter particle, then $\left(1+\frac{m_i}{m_j}\right)\le2$ and $\left(1+\frac{r_j}{r_i}\right)\ge2$ and we conclude that the condition on the elasticity constant is:
\begin{equation}
    k\gg \frac{4\pi}{3} \rho r_{\max} v_{\max}^2
    \label{eq:condk}
\end{equation}
where $r_{\max}$ is the radius of the biggest particle in the simulation and $v_{\max}$
is the speed of the fastest particle in the simulation.
In most cases, the above condition yields a minimum constant for physical reliability that is much smaller than the actual Young modulus of rocks, which is of the order of $10^{11}$~barye ($10^{10}$~N/m$^2$).

The condition on the time step is inextricably linked to the chosen elasticity constant, since a larger constant yields more stiff particles and therefore smaller compressions, which require a smaller time step to handle. The condition we enforce is
\begin{equation}
    v_{\max}\delta{t}\ll\Delta_{ij}
\end{equation}
which, after analogous calculations, yields
\begin{equation}
    \delta{t}\ll\sqrt{\frac{m_{\min}}{k}}
    \label{eq:condt}
\end{equation}
where $m_{\min}$ is the mass of the smallest particle in the simulation. We note that this is equivalent to imposing a time step that is much shorter than the oscillation period of a spring with an elastic constant $k$ loaded with the mass of the lightest particle considered.

It should be noted that in simple binary collisions, the actual value of the elasticity constant used predominantly affects the duration of a particle-particle bounce. It has a negligible impact on the forces involved and on the result of the collision\footnote{The elastic constant, however, being directly related to the material's Young modulus, does affect the speed of sound in the particles}. Depending on the application, it may not be necessary to use the actual elasticity of the material in order to obtain reliable results. For that reason, it is more productive to select a value that satisfies Eq.~\ref{eq:condk} and then use Eq.~\ref{eq:condt} to derive the required time step. In our case, since we are interested in collisions at speed smaller than but close to the speed of sound, the adopted elasticity constant is similar to the physical one.

\subsection{Rubble-pile Formation}
\label{sec:formation}
In astronomy, a rubble pile is an object made of individual particles that have accumulated under the effect of self-gravity, for example after an energetic collision that has shattered a previously unfractured object (e.g., \citealt{Bagatin2001}). In addition, no single particle should contain more than half the total mass of the system. To initialize such a cluster, particles are uniformly seeded in a  spherical volume. A collision/overlap check is run across all particle pairs and if a collision/overlap is detected, one of the two particles is moved randomly to another location within the spherical domain limit. This process is repeated with a tolerance of 200 tries for all particles. If any collisions/overlaps are detected after 200 tries, the sphere radius is increased by the diameter of the largest particle. All particle positions are then randomly re-seeded in the new volume and the process is re-started. This eventually leads to a reasonably well packed cluster of particles (see the left panel of Figure~\ref{fig:formation}). The loosely packed particles are then allowed to collapse under their own gravity into a stable structure. Rubble-pile formation is considered complete when the kinetic energy of the system has reduced to a negligible value compared to the collapsing phase and to the collision energy involved in the subsequent simulation. All rubble piles used were relaxed until their kinetic energy was less than $10^{-6}$ the energy involved in the fastest collision, equivalent to less than $10^{-5}$ times their gravitational binding energy. An example of a fully formed rubble pile is shown in the right panel of Figure~\ref{fig:formation}. Table~\ref{tab:parms} reports a list of symbols and, when relevant, the values that were adopted in our simulations.

We used the procedure described above to assemble asteroids with three sizes, the particle radii ranging from 95 to 203 km. The number of particles in the three was 500, 1250, and 5000.  Each asteroid is made up of particles of three different sizes such that each particle size makes up a third of the overall mass of the asteroid. See also Sect~\ref{sec:results} for further details.

\subsection{Cluster-finding algorithm}

A post-processing code was developed to compute the outcome of the collision in term of clusters of particles with long-term stability. This action is not trivial, since the simulation cannot be carried on within a reasonable time until all particles are either unbound or part of a stable cluster. The cluster finding algorithm works iteratively in three steps. A first step runs an adaptive Gaussian smoothing filter with a local standard deviation equal to ten times each particle radius and identifies the peak of the filtered mass distribution as the center of a potential stable cluster. The second step cycles through each particle that is not already part of the potential cluster deciding whether the particle is a member of the cluster by analyzing whether or not it is in contact with a particle that belongs to the cluster. This step is repeated until no additional contacts are identified. The third step cycles again through each particle that is not already part of the potential cluster deciding whether the particle is energetically bound to the cluster. The particle velocity with respect to the center of mass of the cluster is compared to the escape velocity at the position of the particle\footnote{In this step we adopt a two-body energy criterion, and do not consider the direction of the relative velocity. We also do not investigate whether the particle is on a collision course with another particle. All of these can potentially alter the cluster to which a particle is bound to.} . If the latter is greater, the particle is accepted into the cluster. This step is repeated until no new particles are accepted into the cluster. At this point the whole process is repeated on all the remaining particles that are unbound from all the previously found clusters.  This cycle is repeated until no new cluster with more than a user-defined minimum number of particles is found.
\begin{figure}
    \includegraphics[width=\columnwidth]{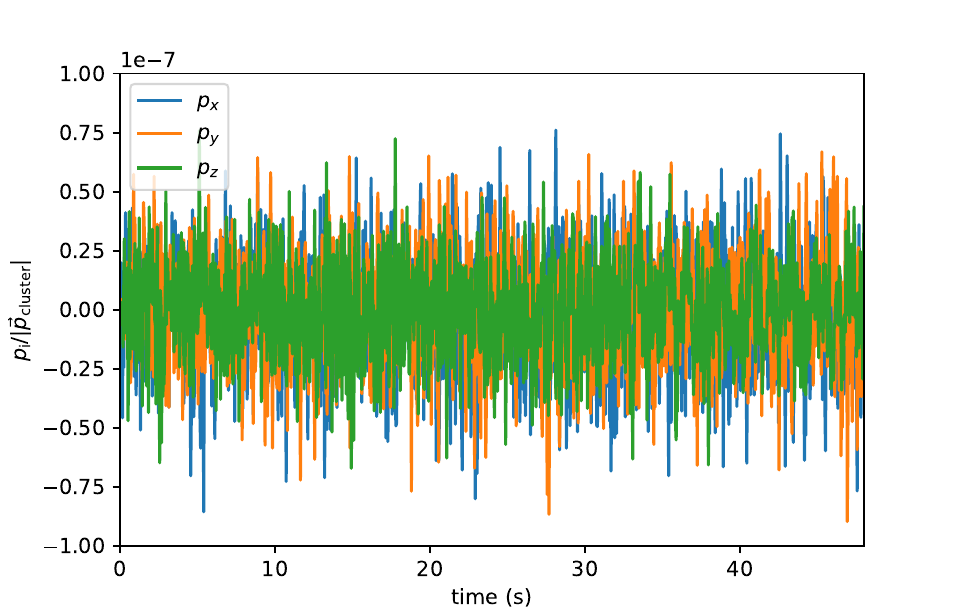}
    \caption{{Momentum conservation test for the DECCO code. The figure shows the result of a $\theta=22^\circ$ high-velocity collision between two large clusters, each with 5000 particles. The total kinetic energy in the collision is 12 times larger than the potential energy of the two clusters. The vertical axis shows the net momentum along each direction normalized by the total momentum carried by one of the clusters.}
    \label{fig:pCons}}
\end{figure}

An example outcome is shown in Figure~\ref{fig:subclusters} where clusters and their member particles are color-coded. The case shown in the figure is the collision of two equal mass rubble piles with three times more translational kinetic than potential energy ($\eta=3$, see below for a formal definition of the parameter $\eta$). In this case six clusters with more than 50 particles are formed in the collision. The outer ones (1-green and 2-blue) are the biggest, with $M\sim4.5\times10^{22}$~g. The inner ones, instead, have smaller mass $M\sim5\times10^{21}$~g. The symmetry in the cluster properties is expected given the symmetry in the initial conditions. It has to be noted that very few of the cases in Figure~\ref{fig:comparaLei} produce more than two clusters in the post-collision configuration. In most cases a single cluster is formed, except for collisions between an identical projectile and target, in which two similar clusters are most likely formed.

\begin{figure}
    \centering
    \includegraphics[width=\columnwidth]{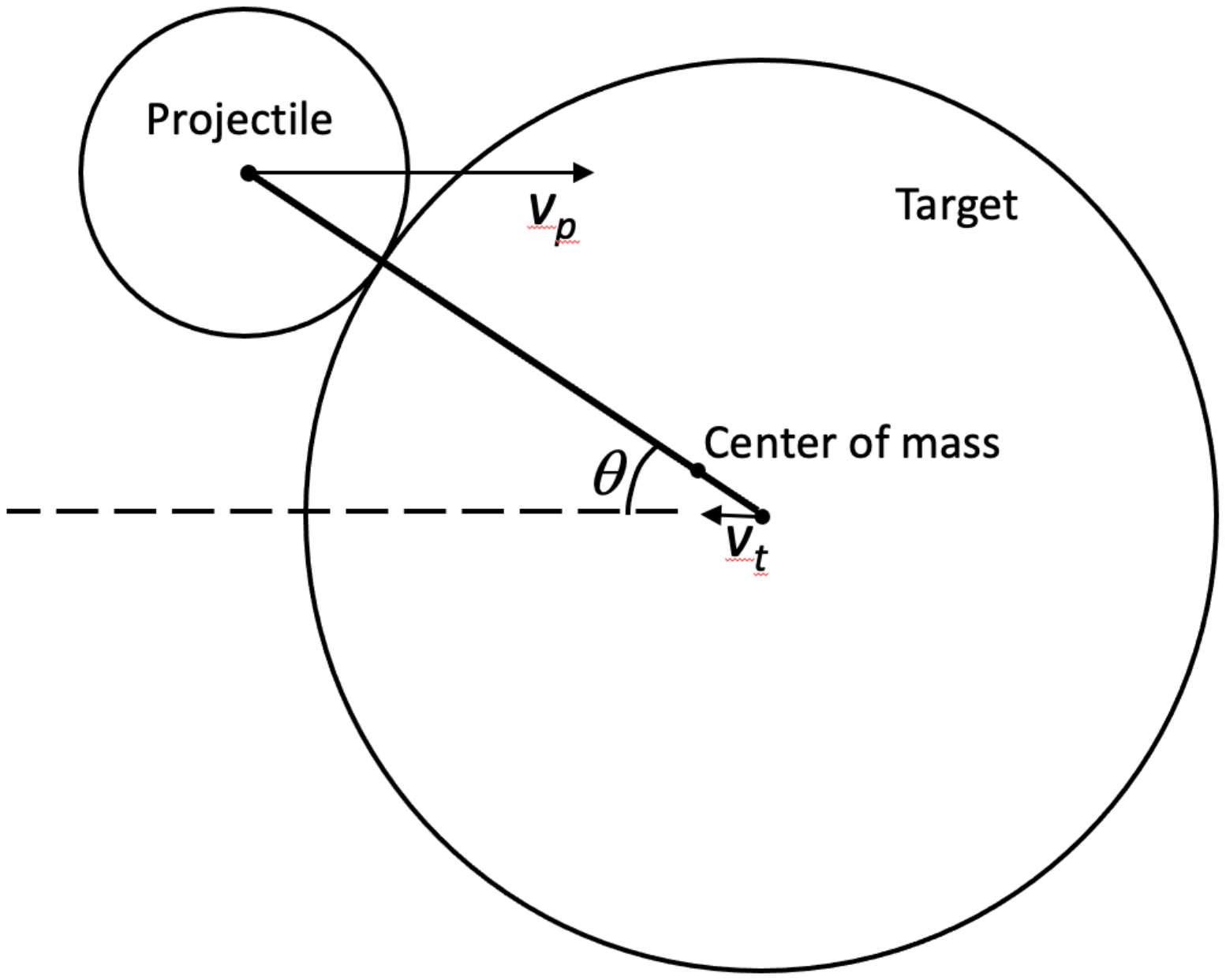}
    \caption{Cartoon showing the collision geometry between a projectile (defined as the object with lower mass) and a target (the object with higher mass). Collisions will be parameterized on their angle $\theta$. All collisions are carried out in the center of mass reference frame in which the velocities of the two objects are anti-parallel by construction.}
    \label{fig:cartoon}
\end{figure}

% \begin{figure*}
% \includegraphics[width=\textwidth]{DART3_v2.pdf}
% \caption{{Post-impact 3D rendering of the Dimorphos model. The impact took place on the right side and small particle ejecta are clearly visible as a result. The bigger ejected particle at the lower right is DART itself, which bounces off the surface at a large angle in our simulation. The right panel shows a zoomed out version of the impact area in the dashed box.}
% \label{fig:DART}}
% \end{figure*}

\section{Code testing and validation}
\label{sec:results}

An initial testing of the code was carried out to ensure that 
basic physics is preserved by testing for conservation of energy (in the case of unity coefficient of restitution) and momentum (for both unity and non-unity coefficient of restitution). All such tests were successful. We show in Figure~\ref{fig:pCons} the result for momentum conservation in an oblique collision ($\theta=22^\circ$, as defined in Figure~\ref{fig:cartoon}). The collision was carried out at the highest velocity (12 times more kinetic than potential energy) between two large clusters (5000 particles each). Figure~\ref{fig:pCons} shows that momentum is conserved in all three spatial components to better than in one part per million. 

In addition to conservation tests, we run a set of validation collisions, in which we aim at replicating the initial conditions from \cite{Leinhardt2012}, repeating their collision simulations, and eventually comparing our results with theirs. We point out, however, that the code used by \cite{Leinhardt2012} is a hard-sphere code, while DECCO uses a soft-sphere method. In addition, \cite{Leinhardt2012} adopts a much smaller coefficient of restitution (0.5) compared to our value (0.9).

\subsection{Validation Runs}
\label{sec:validation}

We test the validity of our code by carrying out a set of simulated collisions between a pair of rubble-piles  of different mass and colliding at different angles. These simulations are designed to emulate the analogous ones carried out by \cite{Leinhardt2012}. We hold the target rubble pile mass (the largest of the pair) fixed, and carry out collisions with projectiles with mass one tenth of the target, one quarter of the target, and equal to the target. For each rubble pile pair, we carry out collisions with angles $\theta=0^\circ$, $22^\circ$, and $44^\circ$, as defined in Figure~\ref{fig:cartoon}. For each rubble pile pair we also carry out collisions with different relative velocity, parameterized by the ratio of the translational kinetic energy involved over the binding potential energy of the two rubble piles. This parameter, which we call $\eta$, is defined as:
\begin{equation}
    \eta=\frac{KE}{|U|}=\frac{\frac{1}{2} m_p v_p^2+\frac{1}{2} m_t v_t^2}
    {\sum_{i<j}G\frac{m_i m_j}{r_{ij}}}
    \label{eq:eta}
\end{equation}
where $m_p$ and $m_t$ are the masses of the projectile and target, respectively, $v_p$ and $v_t$ are their velocities in the center of momentum frame, and $m_i$ and $m_j$ are the masses of particles $i$ and $j$. The parameter $\eta$ is allowed to vary between $0.5\le\eta\le12$. 

All simulations were carried out with rubble piles with a particle distribution comprising three sizes: 7.5, 15, and 22.5 km. The total mass in each particle population was equal, resulting in most of the particles being of the smaller size. All particles had density $\rho=2.7$~g/cm$^3$ \citep{Flynn1999}. The largest (target) rubble pile had 5000 total particles, while the two smaller projectiles had 1250 and 500 total particles, respectively. These choices resulted in rubble piles with mass and radius $(m_{\rm{RP}},r_{\rm{RP}})=(6\times10^{21},95)$, $(10^{22},129)$, and $(6\times10^{22},203)$, in units of g and km, respectively. Impact speed ranged between 0.26 km/s for the slowest equal-mass collision to $\sim 1$~km/s for the fastest collision with the light projectile. These impact velocities are smaller than the speed of sound in dense materials, which is of the order of a few km/s.  However, they may exceed the speed of a wave's propagation through a granular medium at low confining pressure \citep{Sanchez2022}. In addition, some particles might break upon impact, a capability that is not yet implemented in DECCO. All our collision velocities $v_{\rm{coll}}$ exceeded the system escape velocity $v_{\rm{esc}}$, ranging from $v_{\rm{coll}}\simeq\,14v_{\rm{esc}}$ in the slowest case to $v_{\rm{coll}}\simeq250\,v_{\rm{esc}}$.

We adopted a time step $\delta{t}=0.04$~s and elastic constant $k_{\rm{in}}=10^{18}$, with a coefficient of restitution\footnote{This is a somewhat high value adapted from the measurement of \cite{Durda2011}. They found a coefficient of restitution of 0.83 that we round up conservatively to 0.9.} $\epsilon=0.9$. The elastic constant was chosen to respect Eq.~\ref{eq:condk} for the highest energy collisions and kept constant among different simulations (even if the maximum velocity was reduced) for consistency. The time interval also comfortably satisfies the constraint set by Eq.~\ref{eq:condt}. The friction coefficients that we used were $\mu_k=0.3$ and $\mu_r=10^{-3}$.

As indicated above, there are some significant differences between our calculations and the ones presented in \cite{Leinhardt2012}. First, the version of PKDGRAV used in \cite{Leinhardt2012} is based on collisions between hard spheres, not soft spheres like in DECCO. Second, the coefficient of restitution that they adopt is significantly lower (0.5) than the value that we use. Finally, we used rubble piles with a distribution of particle sizes, while \cite{Leinhardt2012} use mono-dispersed particles in their experiments.  Taking these caveats into account, we compare the mass of the largest fragment in our simulation with that reported by \cite{Leinhardt2012} to assess the reliability of DECCO.

Figure~\ref{fig:comparaLei} shows the result of the validation runs. Each panel shows a rubble pile pair of a given mass ratio colliding at a given angle, and multiple relative velocities are shown. Even though subtle differences can be seen, it is clear that the two codes have similar outcomes, even in the medium-velocity regime when the fragment mass is different from either the projectile or the target. In addition, the two codes agree on the location of the sudden change in largest fragment mass that are seen in the intermediate and large angle, equal mass cases. We did not attempt to replicate their results in the high-velocity regime since our code does not include the hydrodynamic step that is used to model elastic waves and fragmentation at supersonic speed.

While other authors have carried out similar calculations with different codes, we were not able to derive all the needed parameters and outcomes from their publications, and a direct comparison was not possible. 
Based on the results of this comparison, we conclude that DECCO successfully compares with previously established results and is safe to use in the subsonic regime. 

\section{Results}
\subsection{Particle size distribution}

\begin{figure*}
    \includegraphics[width=\textwidth]{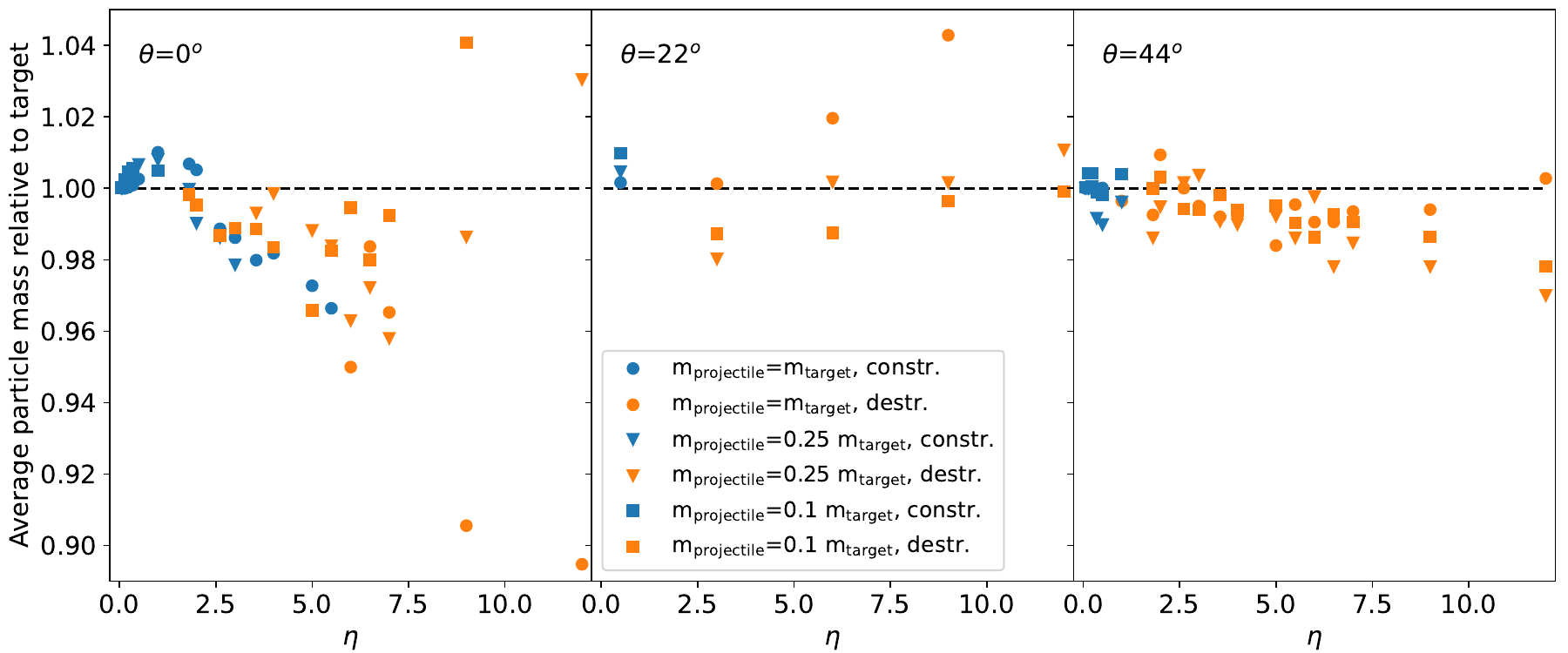}
    \caption{Effect of collisions on the average particle mass of the largest 
    fragment. Blue symbols show cases with constructive collisions ($m_{\rm{lf}}>m_{\rm{t}}$) while orange symbols show the result of destructive collisions. Different symbols are used for systems with different projectile to target size, while the three panels show collisions with increasing collision angle.
    \label{fig:partDistr}}
\end{figure*}

Having verified that the DECCO code respects basic conservation laws and compares well with the result of other codes based on analogous physics, we use the specific capability of DECCO to handle particles of different sizes to investigate the role of collisions in changing the particle distributions of rubble piles. In other words, we study whether the collisions predominantly eject the largest or the smallest particles in the system. This analysis is based on the same set of simulations that were used for the validation comparison of Fig.~\ref{fig:comparaLei}. The results of the analysis are shown in Figure~\ref{fig:partDistr}. While there is significant scattering, there are also interesting patterns. First, any effect is small, at the level of a few per cent. Repeated collisions, however, could result on a build up making the effect more noticeable. Second, most constructive collisions (blue colored symbols) and low velocity collisions tend to preferentially eject small particles, leaving a debris characterized by a bigger particle population. This is especially true for asymmetric collisions (small projectile, squares), for which all constructive collisions result in a fragment with bigger average particle size. Most destructive collisions, instead, tend to eject bigger particles preferentially, with some notable exceptions at very high velocities.

\begin{figure*}
    \includegraphics[width=\textwidth]{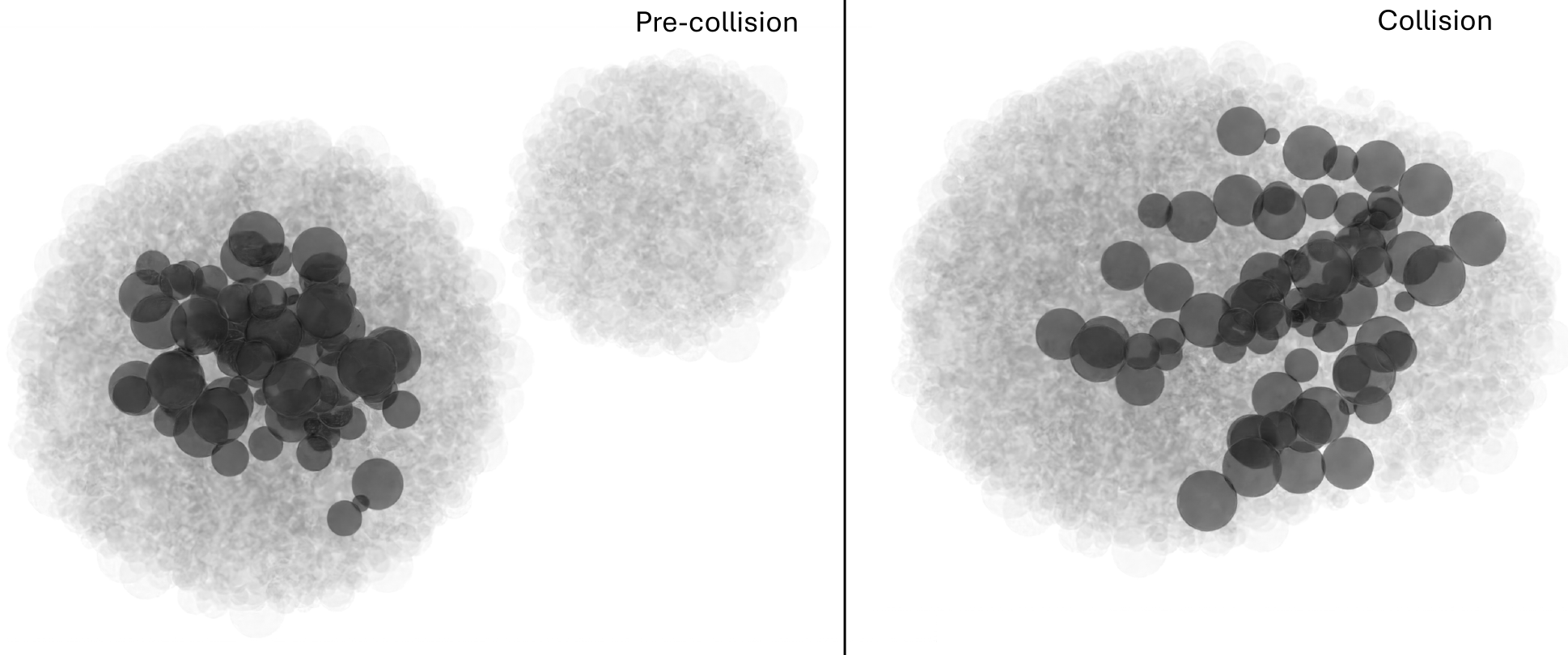}
    \caption{Volume rendering of the  rubble piles before and during their collision. All particles are shown as a gray haze. Particles subject to the highest amount of stress are highlighted in dark gray. The collision is a constructive, low-velocity simulation ($\eta=3$) for a mass ratio 0.25 system with an impact angle of 22 degrees. 
    The stresses are concentrated in the center of the target before the collision but the impact forces transfer the stresses to the area where the impact is happening, creating high stress chains that are carried predominantly by the bigger particles.
    \label{fig:forceChains}}
\end{figure*}

These results can be interpreted as follows: if the energy involved in the collision is not large, it is unlikely that a single, massive particle receives enough extra energy to become unbound from the system. For this reason, constructive collisions tend to eject small particle debris and result in a stable fragment with larger average particle size. At higher energies, instead, the fact that the particles carrying the force chains are predominantly large (see Figure~\ref{fig:forceChains}) creates a setup in which large particles can receive the bulk of the energy of the impact. 

\begin{figure*}
    \includegraphics[width=\textwidth]{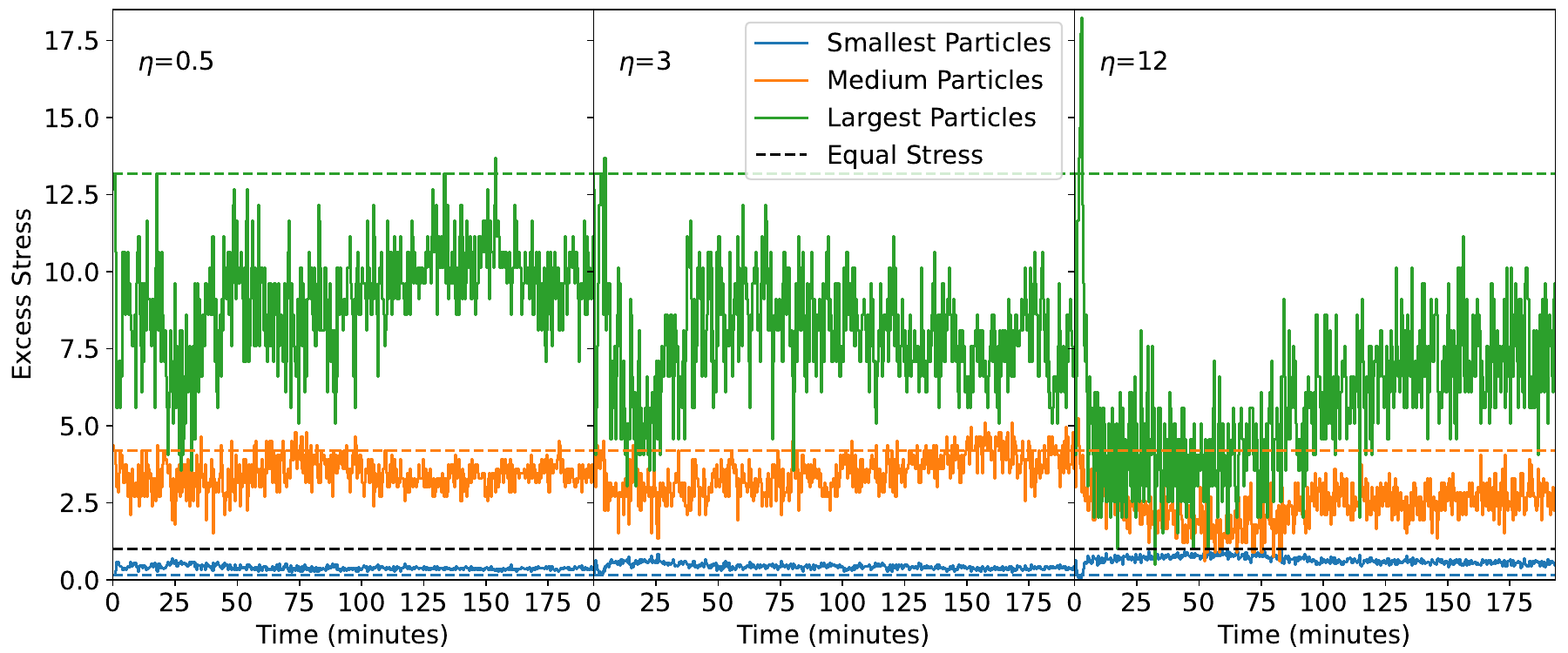}
    \caption{Stress analysis among the different particle sizes for a $\theta=22^\circ$, $m_p/m_t=0.25$ set of collisions of increasing energy (from left to right, $\eta=0.5$, $3.0$, and $12$). The collision takes place, in all cases, at $t=0$. As detailed in the text, values larger than unity denote particle sizes that are subject to an amount of stress that is larger than the average. The figure shows that large particles always carry the largest share of the stress to counter self-gravity in the clusters, and that this imbalance is even more pronounced during high-velocity impacts.  This figure includes  all 6250 particles in the simulation. A similar analysis restricted to only consider particles that remain bound shows comparable qualitative behavior, but the high stress spike in the $\eta=12$ collision at $t\sim5$~s is significantly reduced, indicating that the  particles subject to the highest stress during the collision are eventually ejected.
    \label{fig:strainDistrib}}
    \vspace{.6cm}
\end{figure*}

We further explore this in Figure~\ref{fig:strainDistrib}. The figure is built to explore whether the high stresses are carried equally by particles of all sizes. The data shown are for a $\theta=22^\circ$ collision between a 1250 particle projectile and a 5000 particle target. Three values of the collision energy $\eta$ are shown. To create the figure, we first selected the top 1\%\footnote{Note that the choice of the top 1\% of particles is arbitrary. However, the behavior described in this section is robust against a different choice of what defines the high-stress category of particles.} of particles based on the stress that they are under (computed as the maximum particle overlap with all particles in contact). For each particle size $i$, the y-axis of Figure~\ref{fig:strainDistrib} shows the quantity $100 \, N_{i,1\%}/N_i$, where $N_i$ is the number of particle of size $i$ in the entire simulation, and $N_{i,1\%}$ is the number of particles of size $i$ with the top $1\%$ stress. If all particle sizes were  equally under stress, all y-axis values in the figures would be 1. That is not the case. As a general finding, the largest particles carry most of the stress at all times (the green line is always the largest). In addition, there is a clear evolution during and after the collision. In all cases, the pre-collision stress is emphasized with a horizontal dashed line. Let us first analyze the high-energy collision in the right panel. There is a short, few-minute spike during the collision (also shown in the right panel of Figure~\ref{fig:forceChains}) in which large particles carry most of the  stress, even more than in the pre-collision configuration. This is followed by a post-collision phase (shown in Figure~\ref{fig:strainDistrib} for $5'\lesssim{t}\lesssim100'$) that lasts a few hours during which the rubble pile settles into a new configuration. In this phase, small grains increase their share of stress (green and orange lines are lower, blue line is higher). Eventually, the collision debris settle in a new configuration in which the largest particles still carry most of the stress but to a lesser extent with respect to the initial configuration. A similar pattern is observed for a medium energy collision in the central panel, albeit with reduced amplitude of the oscillations. 

\subsection{Effect of rotation}

\begin{figure}
\centering
\includegraphics[width=1\linewidth]{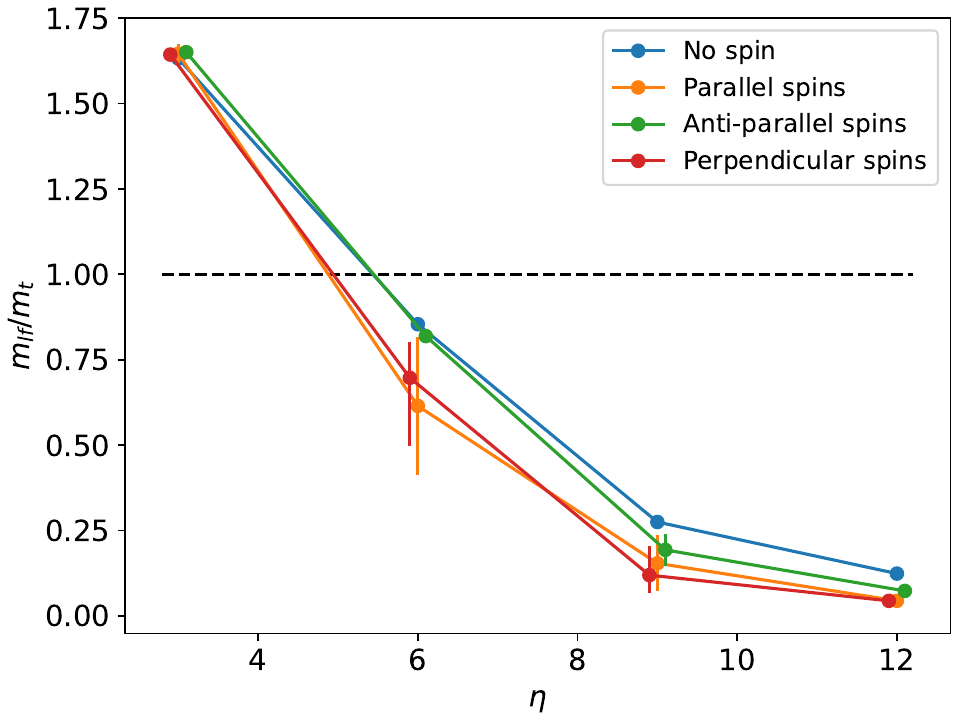}
\caption{Effect of pre-collision spinning on the mass of the largest fragment for head-on collisions ($\theta=0^\circ$). The figure shows the ratio of the mass of the largest fragment over the target mass vs. the energy parameter $\eta$ (defined in Eq.~\ref{eq:eta}). Different spin configurations have been explored and are compared with non-spinning simulations (see legend).}
\label{fig:spinning}
\end{figure}

A new dedicated set of simulations was carried out to investigate the role of rotation on the collision outcome. Introducing rotations opens up a large parameter space in terms of the angular velocity and orientation of the angular momentum vectors for both the projectile and target clusters. To limit the number of simulations, we investigate only the case of head-on ($\theta=0^\circ$) collisions between same-size clusters (5000 particles each). Four different values of the collision energy were investigated ($\eta=3$, 6, 9 and 12). The angular velocity of the clusters was always set up as half of the breakup speed, and the angular momentum vector was either parallel or anti-parallel to one of the three Cartesian axes. The collision velocities are always along the x-axis. The rotational energy of the rubble piles was not included in the calculation of $\eta$, which only considers the translational kinetic energy of the cluster. At half breakout speed, the kinetic energy associated with the cluster rotation would be approximately 10\% of its potential binding energy, and therefore the $\eta$ values of the results in Figure~\ref{fig:spinning} would move to the right by 0.1 for all rotating cases (orange, green, and red).

Figure~\ref{fig:spinning} shows the outcome of the simulations in terms of the mass of the largest fragment. Simulations with similar angular velocity configuration were consolidated in a single data point. For example, a simulation in which the projectile's angular momentum is along the x-axis direction and the target's momentum is along the z-axis direction would fall in the "perpendicular spin" case, as well as projectile's angular momentum along the y-axis and target along the x-axis and so on. In the cases in which more than one simulation is consolidated in a single data point, the error bar shows the dispersion of the various realizations. A particularly relevant case is the one of parallel spins, since planetesimals embedded in an accretion disk are expected to be spun up by the differential rotation of the disk with parallel angular velocities.

Our limited investigation in the effect of spin on the collision outcome suggests that anti-parallel spin has a very small effect, while both parallel and perpendicular spin make the collision more disruptive (the largest remnants' masses being smaller). This could be due to the fact that the surfaces of rubble piles colliding with either perpendicular or parallel spins have a strong sheer velocity, while rubble piles with anti-parallel spins would roll on each other. The latter is, however, true only because the two rubble piles have the same size and angular velocity, in our case. A more thorough investigation of the effect of spin is needed to generalize these results. An analysis of the effect of rotation was carried out by Ballouz and co-authors \citep{Ballouz2014,Ballouz2015}, who investigated the effect of the target rotation in asymmetric collisions (the mass of the projectile being one tenth the mass of the target). Their findings are similar to ours in the fact that faster rotation is associated with smaller fragments (see, e.g., Figure~2 in \citealt{Ballouz2014}).

\section{Summary, discussion, and conclusions}
\label{sec:discussion}

We have presented a new implementation of a soft-sphere, discrete element code for computing the dynamics of self-gravitating rubble piles. The code allows for particles of different sizes, includes inelastic normal forces between particles, gravity, kinetic friction, and rolling friction. Inelastic collisions are implemented by changing the spring constant instantaneously during rebound, ensuring constant coefficient of restitution irrespective of particle size. Short-range forces like Van der Waals forces are not included at this stage, but are expected to be irrelevant at systems size and collision velocities investigated here. As the code is expanded to allow for simulation of small particles (e.g., dust grains) these forces will be included (Kolanz et al., in preparation). Additional limitations of the code are the absence of static and twisting friction, which --- together with the instantaneous change of the spring constant--- limit its applicability only to dynamic calculations, and the fact that particle deformation is not accounted for in calculating surface velocities and torques. This latter simplification is not expected to significantly affect the results as the maximum compression is kept small to ensure stability (see Sect.~\ref{sec:stability}). Post-processing analysis of the presented results has revealed that even in the fastest collisions that we have analyzed ($\eta=12$) the average spring compression is $\sim0.03$ per cent, with less than 1 per cent of the particles ever experiencing a compression value larger than 1 per cent of their radius.

Tests of energy and momentum conservation, as well as validation runs for the code have been successfully carried out to test its reliability by reproducing previous results from an analogous, well-established code (Figure~\ref{fig:comparaLei}; \citealt{Leinhardt2012})\footnote{See, however, Sect.~\ref{sec:validation} for a discussion on the technical and material differences between our simulations and the validation runs they are compared to.}. Given the proven validity of the code we have carried out two experiments. In the first one we have studied the particle size composition of the largest remnant from rubble pile collisions, looking for patterns of change. We found that low-velocity collisions tend to predominantly eject small particles, resulting in remnants with larger average particle size. We note that our input rubble piles do not have a segregated particle distribution with smaller particles on the surface. For this reason, more realistic simulations may find an even larger effect. For large velocity impacts that leave behind small fragments, instead, we find that the average particle size is smaller. We ascribe this behavior to the fact that the larger particles appear to be subject to the largest amount of stress, creating the force chains that propagate the force of the impact through the rubble piles (see Figures~\ref{fig:forceChains} and~\ref{fig:strainDistrib}). It should also be noted, however, that we have only tested one particle size distribution that has few large particles in it, and the role of the distribution on the collision outcome should be investigated to establish the robustness of this result. Studying the stress distribution among particles of different sizes also led us to note that the post-collision configuration of the clusters is not congruent with their initial condition. The smallest particles carry an increased fraction of the stress in the post-collision configuration. This finding emphasizes the fact that a better approach to building initial cluster configurations is by hierarchically colliding smaller clusters rather than by a monolithic collapse as described in Sect.~\ref{sec:formation}. We plan to investigate the effect of repeated collision in a future publication.

In a second experiment we simulated collisions between fast-spinning rubble piles. We investigated the case of two identical rubble piles colliding head on ($\theta=0^\circ$). We found that collisions between rotating rubble piles lead to smaller remnants for most rotation directions. An exception was the case in which the rotation produces the effect of the two rubble piles rolling on each other. This reduced remnant size effect would be relevant in particular for collisions within accretion disks, in which the angular velocity vectors of the projectile and target objects are expected to be aligned.

While DECCO is not the only code able to carry out the simulations described here, it is important to assess the robustness of numerical results by using different codes with their own intrinsic strength and weaknesses. Future plans in term of both science and development involve adding  short-range and cohesion forces to allow for simulations of small particles  and microscopic systems, parallelizing the code to allow for a larger number of particles, and making the code GPU portable to increase efficiency.

\begin{acknowledgments}
We are grateful to the anonymous referees for their numerous constructive comments and suggestions that significantly improved this manuscript. This project was supported by NASA APRA award 80NSSC19K0330.
\end{acknowledgments}
%

%%%%%%%%%%%%%%%%%%%%%%%%%%%%%%%%%%%%%%%%%%%%%%%%%%
%\bibliographystyle{aasjournals}
\bibliography{asteroids}

\begin{thebibliography}{}
\expandafter\ifx\csname natexlab\endcsname\relax\def\natexlab#1{#1}\fi
\providecommand{\url}[1]{\href{#1}{#1}}
\providecommand{\dodoi}[1]{doi:~\href{http://doi.org/#1}{\nolinkurl{#1}}}
\providecommand{\doeprint}[1]{\href{http://ascl.net/#1}{\nolinkurl{http://ascl.net/#1}}}
\providecommand{\doarXiv}[1]{\href{https://arxiv.org/abs/#1}{\nolinkurl{https://arxiv.org/abs/#1}}}

\bibitem[{An \& Tannant(2007)}]{An2007}
An, B., \& Tannant, D.~D. 2007, Computers \& Geosciences, 33, 513,
  \dodoi{https://doi.org/10.1016/j.cageo.2006.07.006}

\bibitem[{{Asphaug} {et~al.}(1998){Asphaug}, {Ostro}, {Hudson}, {Scheeres}, \&
  {Benz}}]{Asphaug1998}
{Asphaug}, E., {Ostro}, S.~J., {Hudson}, R.~S., {Scheeres}, D.~J., \& {Benz},
  W. 1998, \nat, 393, 437, \dodoi{10.1038/30911}

\bibitem[{Bagatin {et~al.}(2001)Bagatin, Petit, \& Farinella}]{Bagatin2001}
Bagatin, A., Petit, J.-M., \& Farinella, P. 2001, Icarus, 149, 198,
  \dodoi{https://doi.org/10.1006/icar.2000.6531}

\bibitem[{Ballouz {et~al.}(2014)Ballouz, Richardson, Michel, \&
  Schwartz}]{Ballouz2014}
Ballouz, R.-L., Richardson, D.~C., Michel, P., \& Schwartz, S.~R. 2014, 789,
  158, \dodoi{10.1088/0004-637X/789/2/158}

\bibitem[{{Ballouz} {et~al.}(2015){Ballouz}, {Richardson}, {Michel},
  {Schwartz}, \& {Yu}}]{Ballouz2015}
{Ballouz}, R.~L., {Richardson}, D.~C., {Michel}, P., {Schwartz}, S.~R., \&
  {Yu}, Y. 2015, \planss, 107, 29, \dodoi{10.1016/j.pss.2014.06.003}

\bibitem[{{Benavidez} {et~al.}(2012){Benavidez}, {Durda}, {Enke}, {Bottke},
  {Nesvorn{\'y}}, {Richardson}, {Asphaug}, \& {Merline}}]{Benavidez2012}
{Benavidez}, P.~G., {Durda}, D.~D., {Enke}, B.~L., {et~al.} 2012, \icarus, 219,
  57, \dodoi{10.1016/j.icarus.2012.01.015}

\bibitem[{{Blum} \& {Wurm}(2008)}]{Blum2008}
{Blum}, J., \& {Wurm}, G. 2008, \araa, 46, 21,
  \dodoi{10.1146/annurev.astro.46.060407.145152}

\bibitem[{{Calzetti} {et~al.}(2000){Calzetti}, {Armus}, {Bohlin}, {Kinney},
  {Koornneef}, \& {Storchi-Bergmann}}]{Calzetti2000}
{Calzetti}, D., {Armus}, L., {Bohlin}, R.~C., {et~al.} 2000, \apj, 533, 682,
  \dodoi{10.1086/308692}

\bibitem[{{Chapman} {et~al.}(1978){Chapman}, {Williams}, \&
  {Hartmann}}]{Chapman1978}
{Chapman}, C.~R., {Williams}, J.~G., \& {Hartmann}, W.~K. 1978, \araa, 16, 33,
  \dodoi{10.1146/annurev.aa.16.090178.000341}

\bibitem[{Cheng {et~al.}(2018)Cheng, Yu, \& Baoyin}]{Cheng2018}
Cheng, B., Yu, Y., \& Baoyin, H. 2018, Phys. Rev. E, 98, 012901,
  \dodoi{10.1103/PhysRevE.98.012901}

\bibitem[{Clark {et~al.}(2012)Clark, Kondic, \& Behringer}]{Clark2012}
Clark, A.~H., Kondic, L., \& Behringer, R.~P. 2012, Phys. Rev. Lett., 109,
  238302, \dodoi{10.1103/PhysRevLett.109.238302}

\bibitem[{Clark {et~al.}(2014)Clark, Petersen, \& Behringer}]{Clark2014}
Clark, A.~H., Petersen, A.~J., \& Behringer, R.~P. 2014, Phys. Rev. E, 89,
  012201, \dodoi{10.1103/PhysRevE.89.012201}

\bibitem[{{Copi} {et~al.}(1995){Copi}, {Schramm}, \& {Turner}}]{Copi1995}
{Copi}, C.~J., {Schramm}, D.~N., \& {Turner}, M.~S. 1995, Science, 267, 192,
  \dodoi{10.1126/science.7809624}

\bibitem[{{Cundall}(1971)}]{Cundall71}
{Cundall}, P.~A. 1971, Proceedings of the Symposium of the International
  Society for Rock Mechanics

\bibitem[{{Cundall} \& {Strack}(1979)}]{Cundall79}
{Cundall}, P.~A., \& {Strack}, O.~D.~L. 1979, Géotechnique,
  \dodoi{10.1680/geot.1979.29.1.47}

\bibitem[{{Davis} {et~al.}(1985){Davis}, {Chapman}, {Weidenschilling}, \&
  {Greenberg}}]{Davis1985}
{Davis}, D.~R., {Chapman}, C.~R., {Weidenschilling}, S.~J., \& {Greenberg}, R.
  1985, \icarus, 62, 30, \dodoi{10.1016/0019-1035(85)90170-8}

\bibitem[{{Dohnanyi}(1969)}]{Dohnanyi1969}
{Dohnanyi}, J.~S. 1969, \jgr, 74, 2531, \dodoi{10.1029/JB074i010p02531}

\bibitem[{{Draine}(2003)}]{Draine2003}
{Draine}, B.~T. 2003, \araa, 41, 241,
  \dodoi{10.1146/annurev.astro.41.011802.094840}

\bibitem[{{Duley} \& {Williams}(1981)}]{Duley1981}
{Duley}, W.~W., \& {Williams}, D.~A. 1981, \mnras, 196, 269,
  \dodoi{10.1093/mnras/196.2.269}

\bibitem[{{Durda} {et~al.}(2004){Durda}, {Bottke}, {Enke}, {Merline},
  {Asphaug}, {Richardson}, \& {Leinhardt}}]{Durda2004}
{Durda}, D.~D., {Bottke}, W.~F., {Enke}, B.~L., {et~al.} 2004, \icarus, 170,
  243, \dodoi{10.1016/j.icarus.2004.04.003}

\bibitem[{Durda {et~al.}(2011)Durda, Movshovitz, Richardson, Asphaug, Morgan,
  Rawlings, \& Vest}]{Durda2011}
Durda, D.~D., Movshovitz, N., Richardson, D.~C., {et~al.} 2011, Icarus, 211,
  849, \dodoi{https://doi.org/10.1016/j.icarus.2010.09.003}

\bibitem[{{Farinella} {et~al.}(1982){Farinella}, {Paolicchi}, \&
  {Zappala}}]{Farinella1982}
{Farinella}, P., {Paolicchi}, P., \& {Zappala}, V. 1982, \icarus, 52, 409,
  \dodoi{10.1016/0019-1035(82)90003-3}

\bibitem[{{Ferrari} \& {Tanga}(2020)}]{Ferrari2020}
{Ferrari}, F., \& {Tanga}, P. 2020, \icarus, 350, 113871,
  \dodoi{10.1016/j.icarus.2020.113871}

\bibitem[{{Flynn} {et~al.}(1999){Flynn}, {Moore}, \& {Kl{\"o}ck}}]{Flynn1999}
{Flynn}, G.~J., {Moore}, L.~B., \& {Kl{\"o}ck}, W. 1999, \icarus, 142, 97,
  \dodoi{10.1006/icar.1999.6210}

\bibitem[{{Goldreich} \& {Ward}(1973)}]{Goldreich1973}
{Goldreich}, P., \& {Ward}, W.~R. 1973, \apj, 183, 1051, \dodoi{10.1086/152291}

\bibitem[{{Hart} {et~al.}(1988){Hart}, {Cundall}, \& {Lemos}}]{Hart1988}
{Hart}, R., {Cundall}, P.~A., \& {Lemos}, J. 1988, International Journal of
  Rock Mechanics and Mining Sciences and Geomechanics Abstracts, 25, 117,
  \dodoi{10.1016/0148-9062(88)92294-2}

\bibitem[{{Hasegawa} \& {Herbst}(1993)}]{Hasegawa1993}
{Hasegawa}, T.~I., \& {Herbst}, E. 1993, \mnras, 261, 83,
  \dodoi{10.1093/mnras/261.1.83}

\bibitem[{{Herrmann} \& {Luding}(1998)}]{Herrmann1998}
{Herrmann}, H., \& {Luding}, S. 1998, Continuum Mech Thermodyn, 10, 189,
  \dodoi{10.1007/s001610050089}

\bibitem[{{Hestroffer} {et~al.}(2019){Hestroffer}, {S{\'a}nchez}, {Staron},
  {Bagatin}, {Eggl}, {Losert}, {Murdoch}, {Opsomer}, {Radjai}, {Richardson},
  {Salazar}, {Scheeres}, {Schwartz}, {Taberlet}, \& {Yano}}]{Hestroffer2019}
{Hestroffer}, D., {S{\'a}nchez}, P., {Staron}, L., {et~al.} 2019, \aapr, 27, 6,
  \dodoi{10.1007/s00159-019-0117-5}

\bibitem[{{Johansen} {et~al.}(2007){Johansen}, {Oishi}, {Mac Low}, {Klahr},
  {Henning}, \& {Youdin}}]{Johansen2007}
{Johansen}, A., {Oishi}, J.~S., {Mac Low}, M.-M., {et~al.} 2007, \nat, 448,
  1022, \dodoi{10.1038/nature06086}

\bibitem[{{Landau} \& {Lifshitz}(1959)}]{Landau1959}
{Landau}, L.~D., \& {Lifshitz}, E.~M. 1959, {Theory of elasticity}
  (Addison-Wesley)

\bibitem[{{Leinhardt} \& {Stewart}(2012)}]{Leinhardt2012}
{Leinhardt}, Z.~M., \& {Stewart}, S.~T. 2012, \apj, 745, 79,
  \dodoi{10.1088/0004-637X/745/1/79}

\bibitem[{{Lissauer}(1993)}]{Lissauer1993}
{Lissauer}, J.~J. 1993, \araa, 31, 129,
  \dodoi{10.1146/annurev.aa.31.090193.001021}

\bibitem[{{Love} \& {Ahrens}(1996)}]{Love1996}
{Love}, S.~G., \& {Ahrens}, T.~J. 1996, \icarus, 124, 141,
  \dodoi{10.1006/icar.1996.0195}

\bibitem[{Luding(1998)}]{Luding1998}
Luding, S. 1998, in NATO ASI Series, Vol. 350, Physics of Dry Granular Media,
  ed. H.~J. {Herrmann} \& S.~{Luding},
  \dodoi{https://doi.org/10.1007/978-94-017-2653-5_20}

\bibitem[{Luding(2007)}]{Luding2007}
Luding, S. 2007, in IUTAM Bookseries, Vol.~1, IUTAM Symposium on Multiscale
  Problems in Multibody System Contacts, ed. P.~{Eberhard},
  \dodoi{https://doi.org/10.1007/978-1-4020-5981-0_14}

\bibitem[{Lunding(2008)}]{Lunding2008}
Lunding, P. 2008, Granular Matter, 10, 235, \dodoi{10.1007/s10035-008-0099-x}

\bibitem[{{Mahmood} \& {Elektorowicz}(2016)}]{Mahmood2016}
{Mahmood}, A.~A., \& {Elektorowicz}, M. 2016, IOP Conf. Ser.: Mater. Sci. Eng,
  136, \dodoi{10.1088/1757-899X/136/1/012034}

\bibitem[{{Mathis}(1990)}]{Mathis1990}
{Mathis}, J.~S. 1990, \araa, 28, 37,
  \dodoi{10.1146/annurev.aa.28.090190.000345}

\bibitem[{{Michel} {et~al.}(2001){Michel}, {Benz}, {Tanga}, \&
  {Richardson}}]{Michel2001}
{Michel}, P., {Benz}, W., {Tanga}, P., \& {Richardson}, D.~C. 2001, Science,
  294, 1696, \dodoi{10.1126/science.1065189}

\bibitem[{{Michel} {et~al.}(2020){Michel}, {Ballouz}, {Barnouin}, {Jutzi},
  {Walsh}, {May}, {Manzoni}, {Richardson}, {Schwartz}, {Sugita}, {Watanabe},
  {Miyamoto}, {Hirabayashi}, {Bottke}, {Connolly}, {Yoshikawa}, \&
  {Lauretta}}]{Michel2020}
{Michel}, P., {Ballouz}, R.~L., {Barnouin}, O.~S., {et~al.} 2020, Nature
  Communications, 11, 2655, \dodoi{10.1038/s41467-020-16433-z}

\bibitem[{{Murdoch} {et~al.}(2015){Murdoch}, {S{\'a}nchez}, {Schwartz}, \&
  {Miyamoto}}]{Murdoch2015}
{Murdoch}, N., {S{\'a}nchez}, P., {Schwartz}, S.~R., \& {Miyamoto}, H. 2015, in
  Asteroids IV, ed. P.~{Michel}, F.~E. {DeMeo}, \& W.~F. {Bottke}, 767--792,
  \dodoi{10.2458/azu_uapress_9780816532131-ch039}

\bibitem[{{P{\'e}roux} \& {Howk}(2020)}]{Peroux2020}
{P{\'e}roux}, C., \& {Howk}, J.~C. 2020, \araa, 58, 363,
  \dodoi{10.1146/annurev-astro-021820-120014}

\bibitem[{{Radjai} \& {Dubois}(2013)}]{Radjai2013}
{Radjai}, F., \& {Dubois}, F. 2013, {Discrete-element Modeling of Granular
  Materials} (Wiley)

\bibitem[{{Richardson} {et~al.}(2009){Richardson}, {Michel}, {Walsh}, \&
  {Flynn}}]{Richardson2009}
{Richardson}, D.~C., {Michel}, P., {Walsh}, K.~J., \& {Flynn}, K.~W. 2009,
  \planss, 57, 183, \dodoi{10.1016/j.pss.2008.04.015}

\bibitem[{{S{\'a}nchez} \& {Scheeres}(2012)}]{Sanchez2012}
{S{\'a}nchez}, D.~P., \& {Scheeres}, D.~J. 2012, \icarus, 218, 876,
  \dodoi{10.1016/j.icarus.2012.01.014}

\bibitem[{{S{\'a}nchez}(2016)}]{Sanchez2016}
{S{\'a}nchez}, P. 2016, in IAU Symposium, Vol. 318, Asteroids: New
  Observations, New Models, ed. S.~R. {Chesley}, A.~{Morbidelli}, R.~{Jedicke},
  \& D.~{Farnocchia}, 111--121, \dodoi{10.1017/S1743921315008583}

\bibitem[{{S{\'a}nchez} \& {Scheeres}(2011)}]{Sanchez2011}
{S{\'a}nchez}, P., \& {Scheeres}, D.~J. 2011, \apj, 727, 120,
  \dodoi{10.1088/0004-637X/727/2/120}

\bibitem[{{S{\'a}nchez} \& {Scheeres}(2014)}]{Sanchez2014}
---. 2014, Meteoritics \& Planetary Science, 49, 788,
  \dodoi{10.1111/maps.12293}

\bibitem[{{S{\'a}nchez} {et~al.}(2022){S{\'a}nchez}, {Scheeres}, \&
  {Quillen}}]{Sanchez2022}
{S{\'a}nchez}, P., {Scheeres}, D.~J., \& {Quillen}, A.~C. 2022, The Planetary
  Science Journal, 3, 245, \dodoi{10.3847/PSJ/ac960c}

\bibitem[{Santos {et~al.}(2020)Santos, Bolintineanu, Grest, Lechman, Plimpton,
  Srivastava, \& Silbert}]{Santos2020}
Santos, A.~P., Bolintineanu, D.~S., Grest, G.~S., {et~al.} 2020, Phys. Rev. E,
  102, 032903, \dodoi{10.1103/PhysRevE.102.032903}

\bibitem[{Schwartz {et~al.}(2012)Schwartz, Richardson, \&
  Michel}]{Schwartz2012}
Schwartz, S., Richardson, D., \& Michel, P. 2012, Granular Matter, 14, 363,
  \dodoi{10.1007/s10035-012-0346-z}

\bibitem[{{Spinrad}(1987)}]{Spinard1987}
{Spinrad}, H. 1987, \araa, 25, 231, \dodoi{10.1146/annurev.aa.25.090187.001311}

\bibitem[{{Spitzer}(1978)}]{Spitzer1978}
{Spitzer}, L. 1978, {Physical processes in the interstellar medium},
  \dodoi{10.1002/9783527617722}

\bibitem[{{Tanga} {et~al.}(2009){Tanga}, {Hestroffer}, {Delb{\`o}}, \&
  {Richardson}}]{Tanga2009}
{Tanga}, P., {Hestroffer}, D., {Delb{\`o}}, M., \& {Richardson}, D.~C. 2009,
  \planss, 57, 193, \dodoi{10.1016/j.pss.2008.06.016}

\bibitem[{Verlet(1967)}]{Verlet1967}
Verlet, L. 1967, Phys. Rev., 159, 98, \dodoi{10.1103/PhysRev.159.98}

\bibitem[{{Walsh}(2018)}]{Walsh2018}
{Walsh}, K.~J. 2018, \araa, 56, 593,
  \dodoi{10.1146/annurev-astro-081817-052013}

\bibitem[{{Walton} \& {Braun}(1986)}]{Walton1986}
{Walton}, O.~R., \& {Braun}, R.~L. 1986, Journal of Rheology, 30, 949,
  \dodoi{10.1122/1.549893}

\bibitem[{{Weingartner} \& {Draine}(2001)}]{Weingartner2001}
{Weingartner}, J.~C., \& {Draine}, B.~T. 2001, \apj, 548, 296,
  \dodoi{10.1086/318651}

\bibitem[{{Wetherill}(1967)}]{Wetherill1967}
{Wetherill}, G.~W. 1967, \jgr, 72, 2429, \dodoi{10.1029/JZ072i009p02429}

\bibitem[{{Wyatt}(2008)}]{Wyatt2008}
{Wyatt}, M.~C. 2008, \araa, 46, 339,
  \dodoi{10.1146/annurev.astro.45.051806.110525}

\bibitem[{{Zhou} {et~al.}(2017){Zhou}, {Chu}, \& {Xu}}]{Zhou2017}
{Zhou}, L., {Chu}, X., \& {Xu}, Y. 2017, in European Physical Journal Web of
  Conferences, Vol. 140, European Physical Journal Web of Conferences (EDP),
  05005, \dodoi{10.1051/epjconf/201714005005}

\end{thebibliography}

%%%%%%%%%%%%%%%%%%%%%%%%%%%%%%%%%%%%%%%%%%%%%%%%%%
\end{document}